# Model of horizontal stress in the Aigion10 well (Corinth) calculated from acoustic body waves

André Rousseau

*CNRS-UMS 2567 (OASU)*
*Université Bordeaux 1 - Groupe d'Etude des Ondes en Géosciences*
*351, cours de la Libération, F-33405 Talence cedex*
a.rousseau@geog.u-bordeaux1.fr

**Abstract**

In this paper we try to deduce the in situ stresses from the monopole acoustic waves of the well AIG10 between 689 and 1004 meters in depth (Corinth Golf). This borehole crosses competent sedimentary formations (mainly limestone), and the active Aigion fault between 769 and 780 meters in depth. This study is the application of two methods previously described by the author who shows the relationships between in situ horizontal stresses, and *(i)* the presence or absence of double body waves, *(ii)* the amplitude ratios between S and P waves (Rousseau, 2005a,b).

The full waveforms of this well exhibit two distinct domains separated by the Aigion fault. Within the upper area the three typical waves (P, S and Stoneley) may appear, but the S waves are not numerous, and there is no double body wave, whereas within the lower area there are sometimes double P waves, but no S waves.

From those observations, we conclude that the stress domain is isotropic above the Aigion fault, and anisotropic below, which is consistent with conclusions drawn by authors from other kinds of data. The calculation applied to the characteristics of the double P waves suggests that this horizontal stress anisotropy is very high, up to 190 % with the maximum horizontal stress superior to 100 MPa and the minimum horizontal stress close to zero at 939 meters in depth. The amplitude ratios between the S and P wave indicate a very strong compression - which is isotropic - above the Aigion fault.

The apparently amazing quasi-absence of acoustic S waves at 11 kHz in frequency through competent formations of fast acoustic P wave velocity may be explained with the freedom degree of vibration of shear waves, parameter proposed by the author in a previous paper. This parameter is inversely proportional to compression magnitude. As for the presence of the isotropic stress domain within an intensely seismic region, we suppose that the earthquakes allied to the active Aigion fault permit the stress homogenisation of the area located outside the regional stress.

**Keywords**

Borehole acoustic waveforms, double body waves, body wave amplitudes, in situ horizontal stress, stress anisotropy, Corinth region

**Introduction**

The AIG10 well (Corinth, Greece) goes through sedimentary rocks (see the litholog in Rettenmaier et al., 2004) which were investigated by monopole acoustic probe between 689 and 1004 meters in depth. The litholog indicates the Aigion normal fault between 760 and 765 meters, but this fault influences acoustic waves between 769 and 780 meters according to the depth indications of the sonic log. This fault separates two main formations, which are on the upper side platy limestones laying over cataclastic bands (745-760 meters of the litholog depths) and on the lower side limestones beneath radiolarites (765-770 meters of the litholog depths).

Anisotropy investigation was carried out from vertical azimuthal velocities with dipole sonic data (Prioul et al., 2004), which leads to the conclusion that there is a large difference between both sides of the Aigion fault. Even pore pressure appears different according as it is measured above or beneath this fault (Cornet et al., 2004). This paper shows the calculations of the horizontal stress and its anisotropy carried out from the monopole acoustic body waves with the method described by the author in Rousseau (2005a).

**The different aspects of the full waveforms**
The shapes of the full waveforms recorded in the investigated vertical area are very different relatively to depth. However, on the whole, S waves are amazingly absent, except in the area between 745 and 760 meters in depth (within cataclastic bands), and in the vicinity of 846, 833 and 826 meters. This lack of S waves is unusual in the case of competent formations and rapid P wave velocities that are here faster than 5000 m/s, with a frequency of 11 kHz. This fact has been observed by Prioul et al. (2004) who noted the absence of signal emitted by a dipolar tool over 5 kHz. The velocity of those S monopole waves is 2540 m/s, that is to say the half of the P wave velocity, and corresponds to the S (flexural) wave velocities ranged from 2450 to 3050 m/s in Prioul et al. (2004). This velocity is low for competent formations, which is again unusual.

The Stoneley waves are variable in amplitude and arrival time. Those characteristics may be consecutive to open fractures or breakouts (in the case of amplitude decreasing) or to probable ovalisation of the hole (in the case of variable arrival time).

Figure 1 displays three typical examples of the full waveforms recorded at different depths.
- At 747.83 meters in depth, that is to say above the Aigion fault within platy limestone, we observe the three typical kinds of wave : P, S and Stoneley waves. The body waves P and S are not double.
- At 990.14 meters in depth within limestone below the Aigion fault, P wave alone is present with Stoneley wave. The latter wave may be absent or very weak at other depths.
- At 938.94 meters in depth within limestone, the body wave P is double ($P_1$ and $P_2$), whereas S and Stoneley waves are missing.

In fact, the acoustic waves propagating within the calcareous formation situated below the Aigion fault do not display many double P waves. They are more or less visible and more or less standard between 953 and 932 meters in depth and between 877 and 873 meters. Stoneley waves may be present or absent, but S wave are absent in every case.

From the aspect of the acoustic full waveforms of this well, the Aigion fault separates two different domains. Above this fault, there are never double P waves, and the few S waves are simple (not double). Beneath it, double P waves are well individualized at particular depths, but without S wave.

**Stress model and discussion**
In spite of small numbers of double P waves and the absence of double S waves, we have applied the method described in Rousseau (2005a) in order to appraise the in situ horizontal stress and its anisotropy. This method is based upon the horizontal size of the stress deformed area caused by drilling and located around the vertical hole ; body waves supposed to reflect within this area are the second wave of the double waves, and their velocity and the half time between both waves arrivals provide the size of those areas.

The absence of double P wave above the Aigion fault indicates an isotropic stress domain, whereas beneath this fault the stress domain is anisotropic. That is conform to the conclusions emitted by Prioul et al. (2004) from the velocities of flexural waves and by Cornet et al. (2004) from hydraulic behaviour.

The result of the stress model calculated from the double P waves recorded at 939 meters in depth is represented on Figure 2. The goal is to manage the coincidence between the wavelength of the P wave (the arrows in red on Figure 2) and the stress deformed area size (represented by a distance on the radial axis) inside the 1 Mpa isobar. This stress limit is the difference of stress relatively to the regional horizontal stress and is supposed to be the limit of the P wave reflection inside the stress deformed area (see Rousseau, 2005a). The wavelength taken into consideration is 0.45 meter (frequency : 11 kHz ; velocity : 5000 m/s), and the distance of the reflection of the P wave is $(210.10^{-6} sec/2) \times 5000$ m/s (see Figure 1 at 938.94 meters), that is to say 0.525 meter. In order to obtain a correct coincidence, the horizontal stress anisotropy must be as large as at least 190 % assuming the minimum horizontal stress $Q_2$ close to zero and a strong maximum horizontal stress $Q_1$ superior to 100 MPa.

In that case, the value of $[K_1 (=Q_1/\sigma_v) + K_2 (=Q_2/\sigma_v)] /2$ (with $\sigma_v$ the overburden pressure) is slightly larger than 2, widely superior to the value 1.3 considered as the limit from which we are in a shear domain (Maury, 1992 ; Maury and Guenot, 1988 ; Maury and Idelovici, 1995).

In addition, the amplitudes of the S waves – if they are present - compared to those of the P waves are relevant for estimating the in situ stress. In Rousseau (2005b) the author shows that stress magnitude is inversely proportional to the ratio determined by the difference between the logarithms of S wave magnitudes and the logarithms of P wave magnitudes. This ratio ranges from zero to 3.5, which respectively corresponds to high compression to tension.

Figure 3 shows the logarithms of P, S and Stoneley wave magnitudes, as well as the differences Sw-Pw between the S and P wave magnitudes from 748.28 to 744.93 meters in depth, just above the Aigion fault (remind that there is no visible S wave below this fault). The value of Sw-Pw oscillates around zero, which reveals a very high compression that is here isotropic, since there is no double body wave.

Now, how to explain the lack of acoustic S wave beneath the Aigion fault, as well as elsewhere above this fault ? In Rousseau (2005b), we noticed that the value Sw-Pw represents a kind of freedom degree of the transversal vibration of S waves ; thus, we may deduce that against a very high stress the vibrating capacity of S waves, at acoustical frequencies (10-20 kHz), is so much reduced that those waves are invisible because of their too low amplitudes or even are missing because they cannot propagate.

Prioul et al. (2004) attribute the stress anisotropy beneath the Aigion fault to intrinsic fractures and local bedding. That would mean that the regional stress domain would be isotropic, which does not correspond to an active seismic region. We should rather look for why there is an isotropic high stress domain in this region.

The geodynamical model built by the author (Rousseau, 1992), after taking into account all the geophysical data known at this time of the Mediterranean region, explains the regional horizontal stress by the pressure exerted by the motion of magma within a large vertical chimney. The fast-shear direction of 105° in Prioul et al. (2004), which is consistent with the regional maximum horizontal stress, coincides with the stress direction suggested by the model in the Corinth region. As a consequence of numerous shallow earthquakes in this region, the active Aigion fault would act as a frontier between the regional deep crustal highly anisotropic stress and the superficial stress that would tend to horizontal stress homogenisation.

**Conclusion**

The investigation of the acoustic monopole waveforms of the wellbore AIG10 highlights two major characteristics relevant to in situ stress :

- S waves are scarce and located only above the Aigion fault,
- as a function of the absence or presence of double P waves, the in situ stress is isotropic above the Aigion fault and anisotropic below.

The horizontal stress model calculated from the time passing between the two waves of the double P waves indicates a very high anisotropy. The maximum horizontal stress is strong, superior to 100 MPa. The very weak ratios between the amplitudes of S and P waves within the stress isotropic region suggest a strong compression. Those results back up the author's hypothesis about the inverse proportionality between the so-called freedom degree of vibration of shear waves and compression magnitude.

As for the difference in the kind of stress domain on both sides of the active Aigion fault, the stress anisotropy below is that of the stress in the Corinth region, whereas the isotropic stress above would be the consequence of the homogenisation of stress after earthquakes.

**Acknowledgements**

The author is grateful to F. Cornet (IPG Paris) for providing the sonic logs.

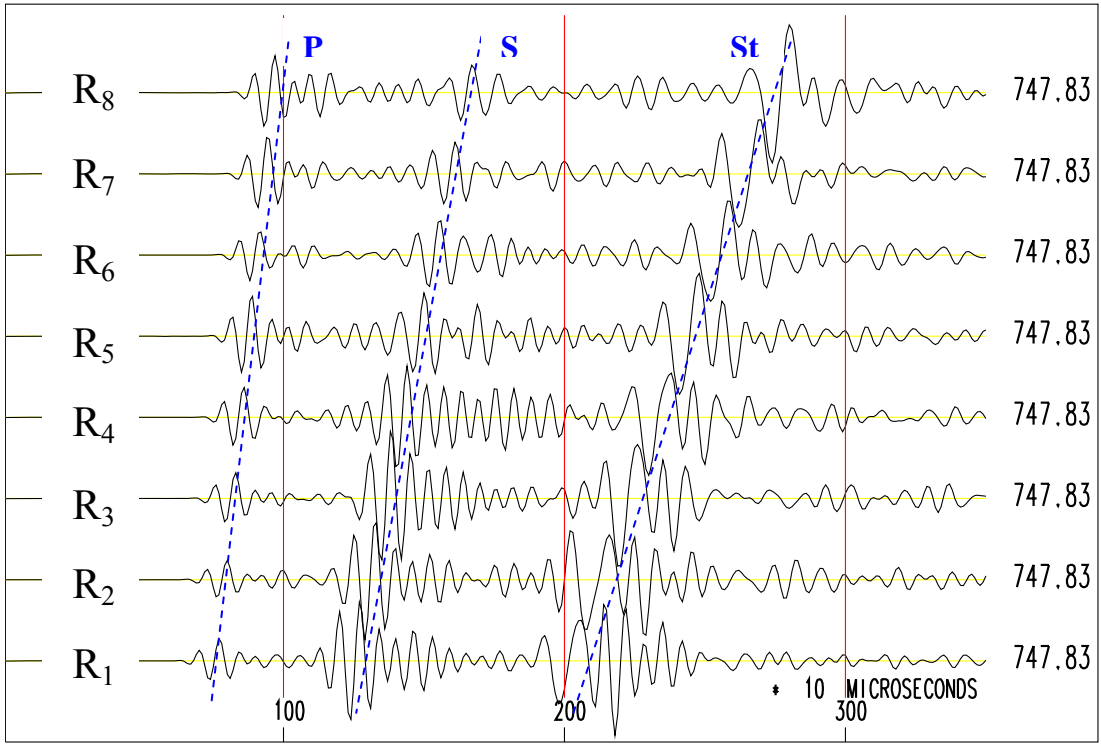

**Figure 1a :** the eight full waveforms ($R_1...R_8$) recorded at 747.83 meters in depth displaying simple P and S waves, and Stoneley (St) wave.
The numbers of x-axis indicate 10 microseconds.

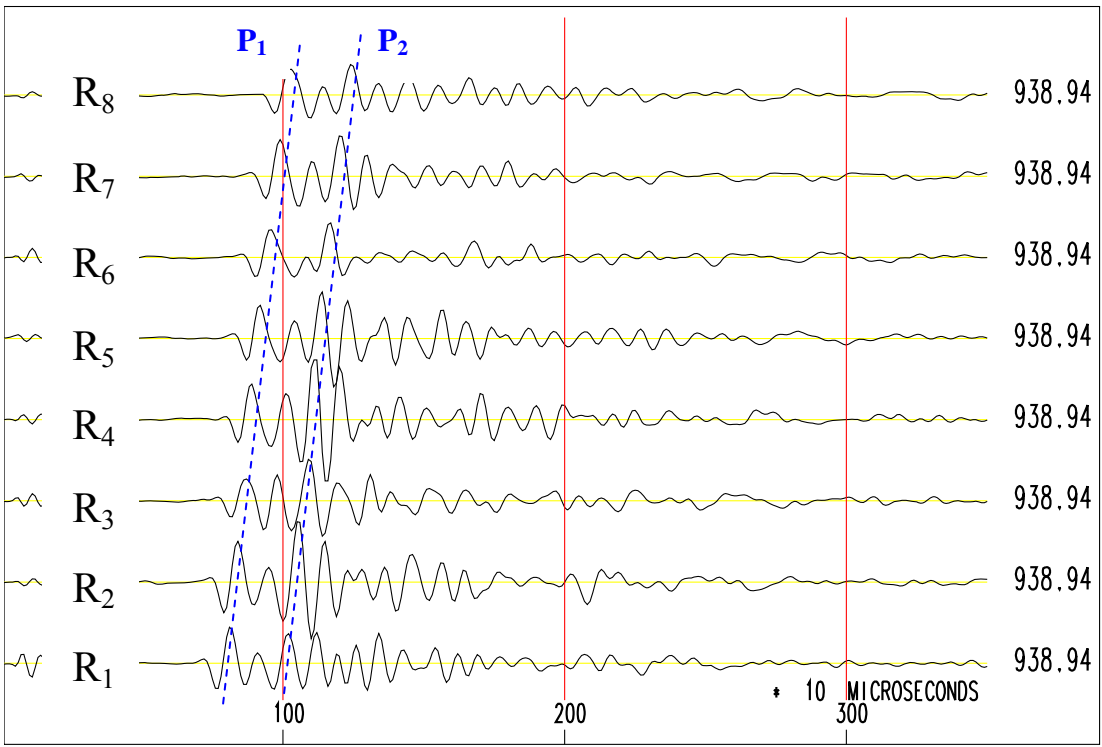

**Figure 1b :** the eight full waveforms ($R_1...R_8$) recorded at 938.94 meters in depth displaying double P wave $P_1$ and $P_2$ alone.
The numbers of x-axis indicate 10 microseconds.

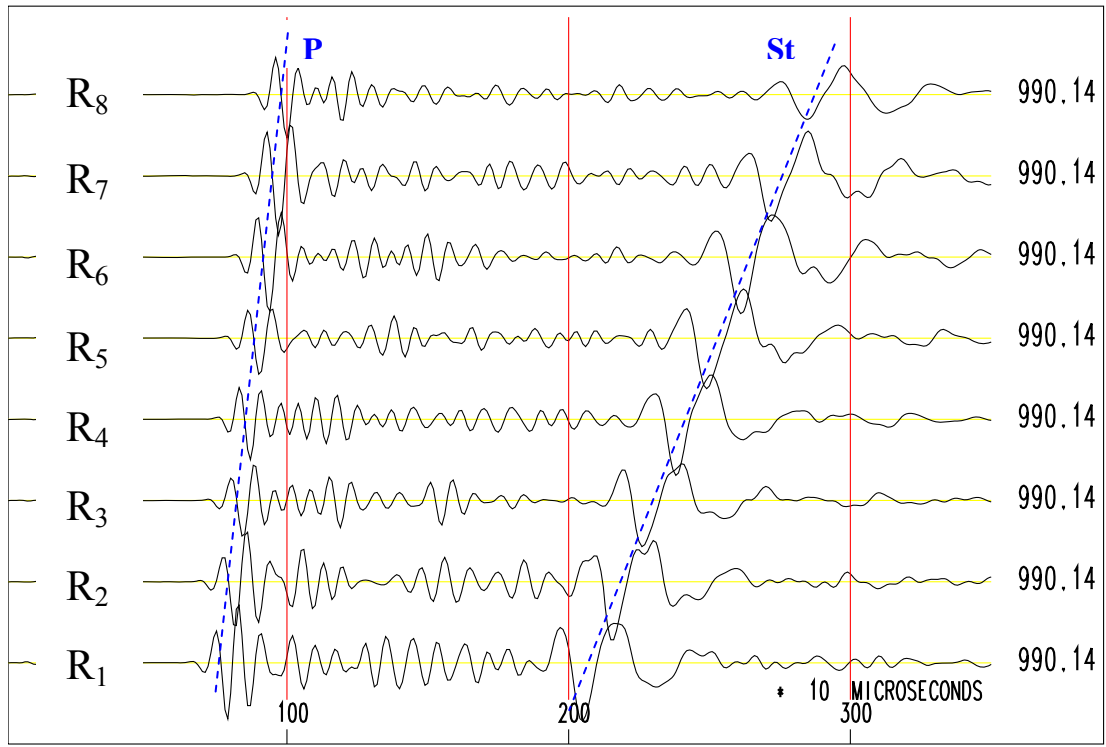

**Figure 1c :** the eight full waveforms (R$_1$…R$_8$) recorded at 990.14 meters in depth displaying simple P wave and Stoneley (St) wave.
The numbers of x-axis indicate 10 microseconds.

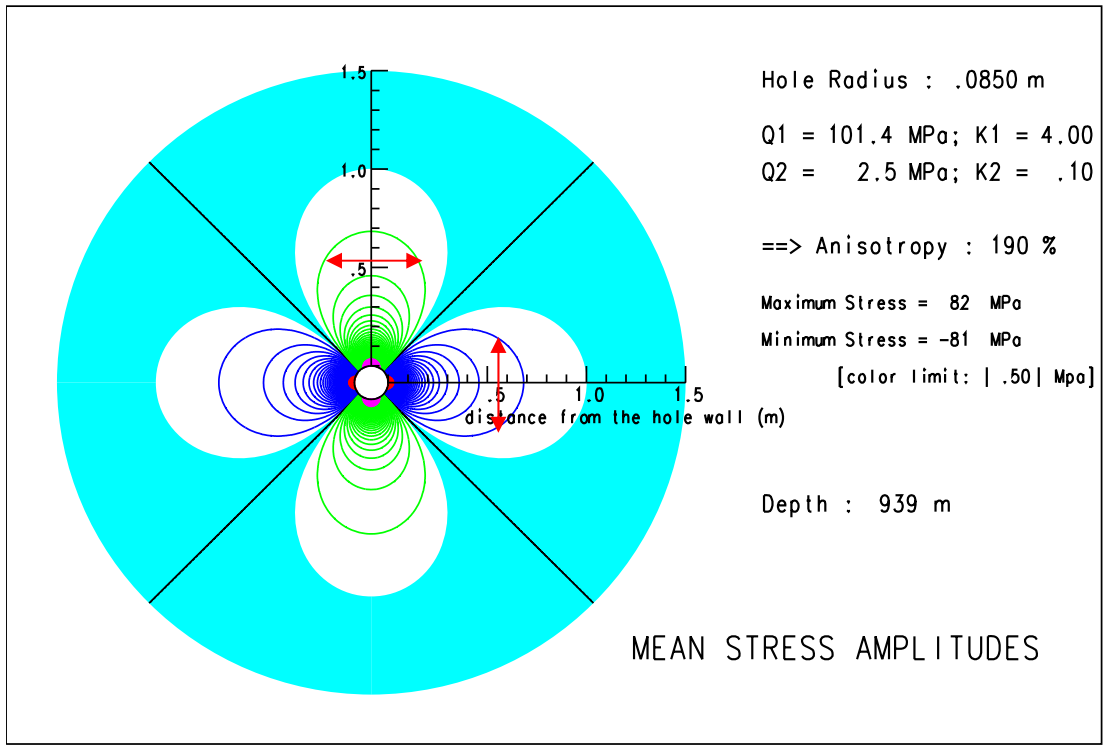

**Figure 2 :** model of stress calculated from the size of the stress deformed areas (see Rousseau, 2005a).
The red arrows represent the wavelength of the P waves.

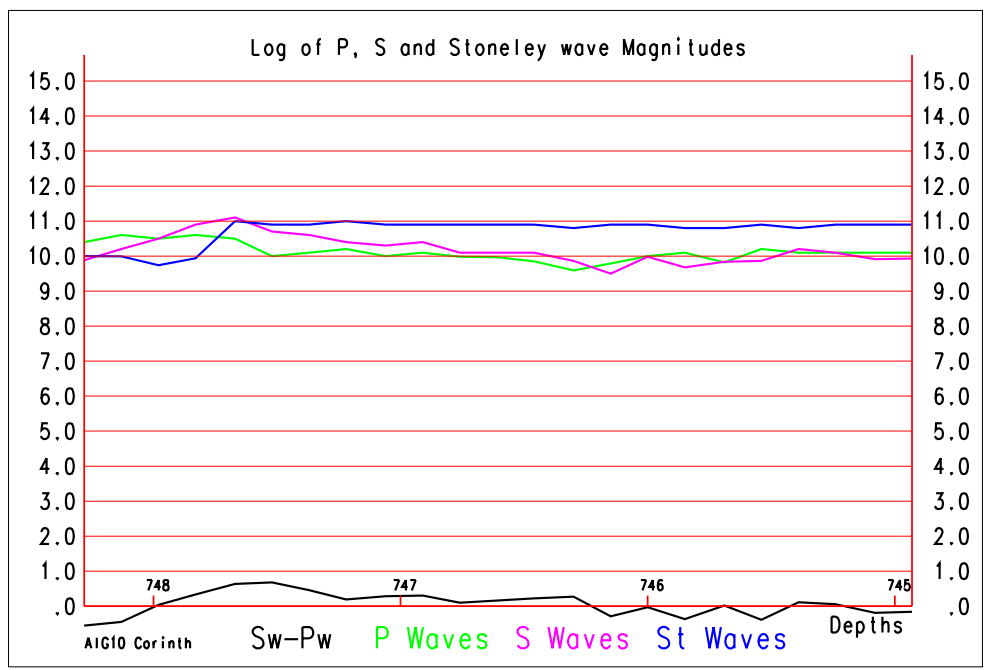

**Figure 3 :** logarithms of the P wave and of the S wave magnitudes with their differences called Sw - Pw.
The logarithms of the Stoneley (St) wave magnitudes are also plotted.

7